\documentclass{article}
\usepackage{mymacros}
\usepackage{axodraw}
\title{Muon Colliders, Monte Carlo and\\Gauge Invariance}
\author{Chris Dams\\Ronald Kleiss}
\let\oldslash\slash
\def\slash#1{\setbox0\hbox{$#1$}\hbox to\wd0{\hss/\hss}\hskip-\wd0\box0}
\def\centerpict#{\vcenter\bgroup\hbox\bgroup\aftergroup\egroup\let\centerpict}
\def\Res{\mathop{\textrm{Res}}\limits}

\begin{document}
\maketitle

\begin{abstract}
\noindent If calculated in the standard way, the cross section for the
collision of two unstable particles turns out to diverge. This is because
part of such a cross section is proportional to the size of the
colliding beams. The effect is called the
``linear beam size effect''. We present a way of including
this linear beam size effect in the usual Monte Carlo integration
procedure. Furthermore we discuss the gauge breaking that
this may cause.
\end{abstract}

\section{Introduction}
The cross section for the collision of two unstable particle generally
diverges. This happens for instance in the Feynman graph
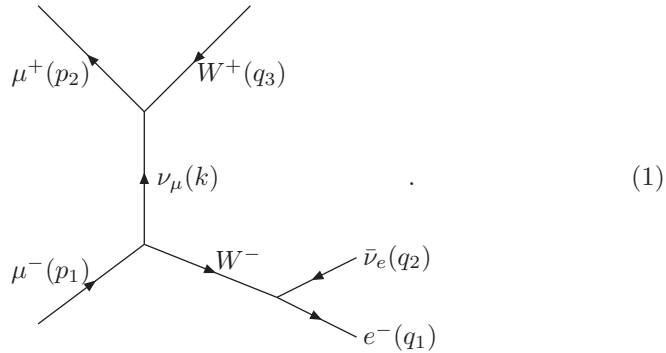
\beq\label{eq:fmdiag}
\centerpict{
\begin{picture}(150,130)
\ArrowLine(10,10)(50,40)\Text(30,25)[rb]{$\mu^-(p_1)$}
\ArrowLine(50,40)(100,20)\Text(77,31)[lb]{$W^-$}
\ArrowLine(100,20)(130,5)\Text(133,5)[l]{$e^-(q_1)$}
\ArrowLine(130,35)(100,20)\Text(133,35)[l]{$\bar\nu_e(q_2)$}
\ArrowLine(50,40)(50,90)\Text(55,65)[l]{$\nu_\mu(k)$}
\ArrowLine(50,90)(10,130)\Text(30,110)[rt]{$\mu^+(p_2)$}
\ArrowLine(90,130)(50,90)\Text(70,110)[lt]{$W^+(q_3)$}
\end{picture}}.
\eeq
The lower half of this graph looks like the decay of a muon. Consequently
the kinematics of the process allows
the momentum~$k$ to be on its mass shell. After all that is what one gets
from the decay of a muon: a muon neutrino on its mass shell.
The factor~$1/(k^2+i\epsilon)$ that occurs in the matrix element
causes a divergence of the total cross section.

In \cite{sing} and \cite{unstable} this problem has been studied in detail,
and it has been shown that this divergence is softened into a finite
peak if the incoming particles are described carefully enough. In this
paper we give a prescription for including this peak in Monte Carlo
simulations. Typically such modifications may result in a violation of
gauge invariance in the amplitude. We study this effect in detail.

\section{Asymptotic States}
In the context of scalar particles, Veltman \cite{veltman} has shown
that the $S$-matrix satisfies unitarity and causality only if one restricts
the in/out states to the stable particles.
Because of this, when considering collisions of unstable particles, we
should use Feynman graphs that take the production process of the
unstable particle into account. We are going to show that we actually
do not need to worry about this as long as the wave packets of the unstable
particles are
much smaller in size than the decay length of the unstable particle.
A complete amplitude for the production
and collision of two muons looks like
\beq
\eqalign{
{\cal A}&=(2\pi)^4i\int \frac{d^4p_1'}{(2\pi)^4}\frac{d^4p_2'}{(2\pi)^4}
	e^{-i\tau_1p_1'\cdot p_1/m_\mu}
	e^{-i\tau_2p_2'\cdot p_2/m_\mu}
	\delta^4(p_1'+p_2'-q_1-q_2)\cr
	&\qquad\psi_{p_2}(p_2')
	\frac{i(-\slash p_2'+m_\mu)}{(p_2')^2-m_\mu^2+im_\mu\Gamma_\mu}
	{\cal M_{\textrm{\scriptsize coll}}}
	\frac{i(\slash p_1'+m_\mu)}{(p_1')^2-m_\mu^2+im_\mu\Gamma_\mu}
	\psi_{p_1}(p_1')\cr
}
\eeq
where
\beq
\psi_{p_1}(p_1')=\int \tilde dp_a'\,\tilde dp_b'\,
	\phi_{p_a}(p_a')\phi_{p_b}(p_b')(2\pi)^4\delta^4(p_a'+p_b'-p_1')
	{\cal M_{\textrm{\scriptsize prod}}}
\eeq
may be viewed as the wave function of the unstable particle.
Notations like $\phi_{p_a}(p_a')$ stand for
the wave function of a particle that
is peaked in momentum space around the value~$p_a$ evaluated at~$p_a'$
The above expression for the wave function of an unstable particles
assumes that the
unstable particle is produced in a two-to-one process. We assume this
only for the sake of simplicity of notations. If there are other
outgoing or incoming particles their wave functions can easily be added.
Also note the factors~$e^{-i\tau_ip_i'\cdot p_i/m_\mu}$. These factors
are translations of the wave function. The point of these translations is
that they make sure that the unstable particles are produced away from the
spot where they collide. The invariant distance that the unstable particle
travels before colliding is~$\tau_i$.

Now we are going to use the assumption that the wave packets are much
smaller than the decay length. This has as a consequence that in momentum
space the wave packets are much broader than the decay length. Because of this
we may assume that they are constant functions of $(p_1')^2$ resp $(p_2')^2$
over a range of several times $m_\mu\Gamma_\mu$.
Therefore
it is possible to integrate the expression for~$\mathcal A$ given above
over the values of~$(p_1')^2$ and $(p_2')^2$. We only have to integrate
the factors contained in the quantity~$F$ that is defined to be given by
\beq
	F=e^{-i\tau_1p_1'\cdot p_1/m_\mu}
	e^{-i\tau_2p_2'\cdot p_2/m_\mu}
	\frac1{[(p_2')^2-m_\mu^2+im_\mu\Gamma_\mu]}
	\frac1{[(p_1')^2-m_\mu^2+im_\mu\Gamma_\mu]}.
\eeq
We integrate along a path parameterized as $p_{1,2}'(t)$.
This parameterization is done according to
\beq
\eqalign{
	p_1'(t)&=p_1'(0)+tc;\cr
	k'(t)&=k'(0)+tc;\cr
	p_2'(t)&=p_2'(0)-tc.\cr
}
\eeq
If we want to integrate over the value of~$(p_1')^2$ we choose $c$ to be a
four vector that is a linear combination of $p_1$,~$p_2$ and $k$ such that
it is orthogonal to the latter two vectors but not to~$p_1$. This
parameterization is chosen such in order to satisfy momentum conservation
and furthermore to be on a constant $k^2$-plane in order not to
get difficulties with the singularity at $k^2=0$. After doing this integral
and an analogous one over the value of $(p_2')^2$, we find the result
\beq
\eqalign{
{\cal A}&=(2\pi)^4i\int \tilde dp_1'\,\tilde dp_2\,
	e^{-i\tau_1p_1'\cdot p_1/m_\mu}e^{-\Gamma_\mu\tau_1/2}
	e^{-i\tau_2p_2'\cdot p_2/m_\mu}e^{-\Gamma_\mu\tau_2/2}\cr
	&\qquad\delta^4(p_1'+p_2'-q_1-q_2)
	\psi_{p_2}(p_2')
	u_i(p_2')\bar u_i(p_2')
	{\cal M_{\textrm{\scriptsize coll}}}
	u_j(p_1')\bar u_j(p_1')
	\psi_{p_1}(p_1').\cr
}
\eeq
This is (except for the decay factors $e^{-\Gamma_\mu\tau_i/2}$)
exactly the same as if we had started with incoming muons
on their mass shell. The conclusion is that if we have wave packets that
are much smaller than the decay length of the unstable particles we may treat
them as if they were asymptotic states.

This has no bearing on the question of gauge invariance.
Matrix elements depend on
the masses of the particles. If masses are chosen such that the muon
is no longer unstable (by assuming the electron to be heavier than the muon,
so that the decay is forbidden), the matrix element is gauge invariant, so it
must also be if masses are chosen accordingly to their measured
values.

\section{The Linear Beam Size Effect}
We observed that the divergence in the cross section is caused by a peak
in the matrix element in momentum space. A sharp peak in momentum space
means a long range effect in position space. Indeed, the decay product of
a decaying muon can travel over arbitrary distances. The consequence
is that the cross section becomes proportional to the size of the beam.
In colliders the longitudinal beam size is much larger than the transverse
one. Consequently, the cross section is actually proportional to the
transverse beam size, to be denoted by~$a$.
The precise definition of this quantity can be found in~\cite{sing}.
In the same paper a more rigorous version of this qualitative argument
was given. In~\cite{unstable} it was shown that the quantities used in
the rigorous argument can replaced by covariant ones.

The part of the cross section proportional to the beam size is given by
\beq
\sigma{\textscr{semi-sing}}=a\pi\int d\sigma_{\textscr{red}}\frac1{|k_\bot|}
	\delta(k^2-m^2),
\eeq
where $\sigma_{\textscr{red}}$ is the cross section with the offending
propagators~$(k^2-m^2\pm i\epsilon)^{-1}$ removed.
$k_\bot$ is by definition given by $k+\alpha p_1+\beta p_2$ with $\alpha$~and
$\beta$ chosen such that $k_\bot\cdot p_{1,2}=0$.

The above formula gives the part of the cross section proportional to
the beam size, but it would be more convenient if the linear beam size effect
could be incorporated in the usual
Monte Carlo integration procedure. This can indeed be done
by doing the substitution
\beq
\frac1{k^2-m^2\pm i\epsilon}\to\frac1{k^2-m^2\pm i|k_\bot|/a}.
\eeq
If we use the approximation
\beq
\frac1{(k^2-m^2)^2+|k_\bot|^2/a^2}\sim\frac{a\pi}{|k_\bot|}\delta(k^2-m^2),
\eeq
these two expressions become equal. This approximation only needs to be
valid around the peak at $k^2=m^2$, which will generally be the case. The
only property that is needed for this to be true is that the reduced cross
section~$d\sigma_{\textscr{red}}$ does not vary much
in $k^2$ at the value~$m^2$ on momentum-squared scales of the order
of~$|k_\bot|/a$. The
contribution of regions of phase space away from this peak can be as large
or larger as the result due to the peak. In~\cite{sing} the matrix element
was split up into a part due to the peak and a part due to the rest of
the phase space to account for this. Our $i|k_\bot|/a$-prescription
gives a good
approximation of the matrix element away from the peak at~$k^2=m^2$, so
it makes a more or less arbitrary split-up of the cross section unnecesary.

\subsection{Gauge Invariance}
The above prescription breaks gauge invariance. We study the process
$\mu^-+\mu^+\to e^-+\bar\nu_e+W^+$. To do this,
six Feynman graphs with $\gamma$, $W^\pm$ and~$Z^0$ as fundamental
bosons are needed. The propagators of the massive bosons must be given
a width. This does affect
the gauge invariance of the amplitude. In~\cite{four} it was shown that
just using the $iM\Gamma$-prescription may lead to
grossly inaccurate results. However, in this paper we
want to focus on the effect of the gauge breaking caused by our
$i|k_\bot|/a$-prescription. For this reason we use the pole scheme for the
massive bosons, so that they do not break the gauge.
What flavour of this scheme we
actually used can be found in appendix~\ref{sec:pole}. It turns out that in
the $R_\xi$-gauge, no gauge
dependence due to the $i|k_\bot|/a$-prescription is found,
although we actually broke gauge invariance.
I.e., the results do not depend on the gauge
parameter~$\xi$. This can be understood from the Feynman graph displayed
in equation~\ref{eq:fmdiag}. The gauge dependence comes in via a term
proportional to $(q_1+q_2)_\mu(q_1+q_2)_\nu$ that occurs in the
$W^-$-propagator. However, this term disappears because one of these
factors~$q_1+q_2$ is to be contracted with the current containing the outgoing
fermions. These are to be taken massless, so consequently this does 
not contribute, regardless of the gauge breaking that may occur at the
other side of the $W^-$-propagator. To see that our prescription actually
breaks gauge invariance we used the axial gauge. In this gauge the
undressed propagator of the $W$-particle is given by
\beq
\Delta(k)_{\nu\mu}
	=\frac{
			-i\left(g_{\mu\nu}
			-\frac{n_\nu k_\mu+n_\mu k_\nu}{n\cdot k}
			+k_\nu k_\mu\frac{n^2}{(n\cdot k)^2}\right)}{
		k^2-M_W^2+i\epsilon}.
\eeq
The expression for the squared matrix element
can be rewritten in such a way that all gauge breaking terms
are proportional to $|k_\bot|/a$ or the square of this quantity.
The axial gauge is not very easy to work with in practice, because one
either has propagators that mix longitudinal gauge bosons with would-be
Goldstone bosons or, if propagators are diagonalized, rather complicated
expressions for the vertices. Details are discussed in~\cite{axial}.
To find the gauge breaking terms in the
unitary gauge we calculate the difference
\beq
|\Mel|^2_{\textscr{gauge-break}}=
	|\Mel|^2_{\textscr{unitary gauge}}-|\Mel|^2_{\textscr{gauge invariant}}.
\eeq
The gauge invariant quantity is calculated by using the axial gauge
and the gauge invariant prescription
\beq
\Mel_{\textscr{gauge invariant}}
	=\frac{\Res_{k^2=m^2}\Mel}{k^2-m^2+i|k_\bot|/a}+\Mel_{\textscr{regular}}
\eeq
that gives a gauge invariant quantity in the spirit of the pole scheme.
This calculation was done in the axial gauge to check that we actually
get a quantity that does not depend on the gauge vector~$n$. The
algebra was done using the C++ computer algebra library GiNaC.
C.f.,~\cite{ginac}.
We find that the quantity~$|\Mel|^2_{\textscr{gauge-break}}$ is, compared
to the rest of the cross section, a factor~$|k_\bot|/(as)\sim 1/(a\sqrt s)$
smaller. Numerically that is a factor $7\cdot 10^{-14}$ for
$\sqrt{s}=150\,\textrm{GeV}$ and $a=\sqrt\pi\cdot 10\,\mu\textrm m$ (which
is a reasonable value).
In ref~\cite{four} it was shown that gauge
breaking effects can get enhanced by a factor as large as~$s/m_e^2$, but
even if this would happen, the gauge breaking due to our handling of the
linear beam size effect remains negligible (note that in the context
of muon colliders one would actually expect a factor $s/m_\mu^2$ for the
case discussed in~\cite{four}).

\section{Conclusions}
The linear beam size effect can be incorporated in the usual Monte Carlo
integration procedure by doing to substitution
\beq
\frac1{k^2-m^2\pm i\epsilon}\to\frac1{k^2-m^2\pm i|k_\bot|/a}
\eeq
in the propagator that causes the divergence. This can be done in a gauge
invariant way, but in the unitary gauge the gauge breaking effect is
so small that it is safe not to worry about the gauge dependence. The
gauge breaking effect of the $iM\Gamma$-prescription is much larger
than that due to the~$i|k_\bot|/a$.

\medskip
\noindent\textbf{Acknowledgement:} We would like to thank prof.\ P. van
Nieuwenhuizen for bringing the problem of gauge dependence to our attention.

\appendix
\section{The Pole Scheme}
\label{sec:pole}
To describe resonances,
as observed from the $W$~and $Z$~particles, one needs a resummed propagator
that contains a factor~$(p^2-M^2+iM\Gamma)^{-1}$.
The problem with this propagator
is that it breaks gauge invariance, which means that observable quantities
depend on the gauge choice. The pole scheme (c.f.,~\cite{stuart}
and~\cite{stuart2})
is one of the ways to
solve this. To use it, we first observe that both the position of the
pole and its residue are gauge invariant quantities.
They must be because they can be determined by experiment.
The consequence is that every matrix element that involves
such a pole can be written as
\beq\label{eq:split}
\Mel=\frac{F(p^2=M^2-iM\Gamma)}{p^2-M^2+iM\Gamma}-\frac{F(p^2)
	-F(p^2=M^2-iM\Gamma)}{p^2-M^2+iM\Gamma}+\Mel_{\textscr{rest}},
\eeq
where the first term is gauge invariant, as are the second and third
together.

In practice things are a bit more involved than sketched in the previous
paragraph. A matrix element generally depends on more that just~$p^2$ and
thus a prescription is needed to tell us what happens to all the other
quantities that occur in the matrix element if we put $p^2$ equal
to~$M^2-iM\Gamma$. We follow the method outlined in~\cite{stuart}.
Our matrix element contains strings of gamma matrices with spinors at
the beginning and end. These are canonicalized to ensure that all
strings of gamma matrices are linearly independent. This means that if we
have, say, a $\slash p$ and a $\slash q$ in some string of gamma matrices,
we can also have the same string of gamma matrices with the $\slash p$
and $\slash q$ interchanged.
We then have to decide which of these two comes in front. The anti-commutation
relations that one has for gamma matrices are then used to do this.
Also the relations $\slash p u(p)=m u(p)$ and $\slash p v(p)=-m v(p)$
are used whenever applicable.

After this has been done, the strings of gamma matrices that remain are
linearly independent. They are said to form a set of independent covariants.
If the matrix element is going to be gauge invariant,
each coefficient of such a string of gamma matrices must separately be
gauge invariant. So equation~\ref{eq:split} is not used for the
full matrix element but actually for the invariant
coefficient that occur in front of the different products of strings of
gamma matrices. In order to do this, it is
also necessary to eliminate one of the outgoing/incoming momenta
by using momentum conservation for the entire matrix element.
All inner products between in- or outgoing momenta in the matrix element
are expressed in a smallest complete set of lorentz invariant variables.
In the case of the outgoing momenta shown in the graph in
equation~\ref{eq:fmdiag} the set consisting of
\beq
\eqalign{
s&=(p_1+p_2)^2;\cr
t&=(p_1-q_1-q_2)^2;\cr
x&=(q_1+q_2)^2;\cr
y&=(p_1+p_2-q_2)^2;\cr
z&=(p_1-q_1)^2,\cr
}
\eeq
can be chosen. If one uses that the squares of incoming and outgoing momenta
are given by the appropriate masses squared, all inner products between
momenta are determined by specifying the variables~$(s,t,x,y,z)$. Now
setting the square of some internal
momentum in some Feynman graph equal to some value is a well-defined
operation, except for some caveats that we discuss next.

\noindent The caveats are
\items{cijfers}{9}
\item If we have outgoing or incoming vector bosons, we should also treat
	inner products of the form $p\cdot\epsilon$ with $\epsilon$
	the polarization vector as linearly independent covariant quantities.
	Some elements in the set of independent covariants
	contain a factor~$p\cdot\epsilon$.
	In the case of the axial gauge this set furthermore includes factors that
	are inner products with the gauge vector~$n$.
\item In the unitary gauge, the inner product of a polarization vector
	with the momentum of the particle to which the polarization vector belongs
	is zero. For this reason, these inner products should not appear in the
	set of independent covariants, nor in the coefficients that multiply
	them. The same applies to the
	inner product of polarization vectors with the gauge vector~$n$ in the
	axial gauge.
\item In the axial gauge the property holds that if we have outgoing or
	incoming vector bosons the matrix element
	becomes zero if the polarization vector of a
	vector boson is replaced by its momentum. It is a feature of the axial
	gauge that this not only happens for massless gauge bosons but also
	for massive ones. This shows that the set
	of covariants that we had is not really linearly independent.
	To see how this can be solved consider a matrix element of the form
	\beq
	\Mel=\epsilon\cdot p_1 F_1+\epsilon\cdot p_2 F_2+\epsilon\cdot p_3 F_3.
	\eeq
	Here the inner products~$\epsilon\cdot p_i$ are the covariants and
	the~$F_i$ are the invariant functions.
	If we know that the relation
	\beq
	q\cdot p_1 F_1+q\cdot p_2 F_2+q\cdot p_3 F_3=0
	\eeq
	holds, we can eliminate $F_1$ from the matrix element. We get
	\beq
	\Mel
		=\left(\epsilon\cdot p_2
							-\epsilon\cdot p_1\frac{q\cdot p_2}{q\cdot p_1}\right)F_2
		+\left(\epsilon\cdot p_3
							-\epsilon\cdot p_1\frac{q\cdot p_3}{q\cdot p_1}\right)F_3.
	\eeq
	Thus we have actually reduced the set of covariants from three to two in
	this example. This boils down to doing the substitution
	\beq
	\epsilon\to \epsilon-\epsilon\cdot p_1\frac{q}{q\cdot p_1}.
	\eeq
	In this substitution the vector~$p_1$ can be chosen to be any linear
	combination of incoming and outgoing momenta. It is advisable to
	choose one that does not yield any singularities in the physical phase
	space due to dividing by~$q\cdot p_1$.
	In the unitary gauge a similar reduction can be carried out. In our
	calculation we chose to get $(q_1+q_2)\cdot q_3$ in the denominator.
	This is equal to $s-x-M_W^2$. This quantity has no poles in the
	physical phase space.
\item One has to be careful about the set of invariant variables. Actually
	the set~$(s,t,x,y,z)$ has a problem. To see this, consider the
	Feynman graph
	\beq
	\centerpict
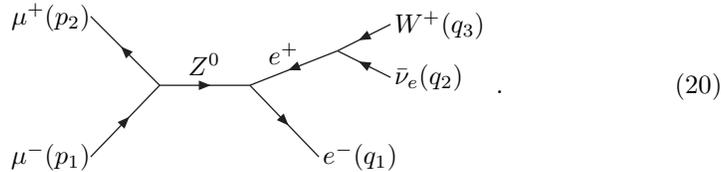
{
		\begin{picture}(170,53)(-13,0)
		\ArrowLine(7,0)(33,27)\Text(7,0)[r]{$\mu^-(p_1)$}
		\ArrowLine(33,27)(7,53)\Text(7,53)[r]{$\mu^+(p_2)$}
		\ArrowLine(33,27)(67,27)\Text(50,31)[b]{$Z^0$}
		\ArrowLine(120,50)(100,40)\Text(122,50)[l]{$W^+(q_3)$}
		\ArrowLine(100,40)(67,27)\Text(80,35)[b]{$e^+$}
		\ArrowLine(67,27)(93,0)\Text(95,0)[l]{$e^-(q_1)$}
		\ArrowLine(120,30)(100,40)\Text(122,30)[l]{$\bar\nu_e(q_2)$}
	\end{picture}
	}.
	\eeq
	The internal electron propagator is given by
	\beq
	S=\frac{-i(\slash q_2+\slash q_3)}{s-x-y+M_W^2}.
	\eeq
	In the pole scheme we should determine the residue for the $Z$-pole.
	This means that we put $s=M_Z^2$ to lowest order. The maximum value
	of $x+y$ is $s$ and occurs in the limit that the outgoing electron
	is produced at rest. We see that the quantity~$1/(s-x-y+M_W^2)$ does
	not have a pole in the physical phase space but if we put $s=M_Z^2$
	it does develop a pole. For this reason we did not use the set
	of parameters~$(s,t,x,y,z)$ but instead the set~$(s,t,\xi,\eta,z)$
	where $\xi=x/s$ and $\eta=y/s$. This set causes no trouble with
	spurious singularities.
\einditems

\noindent The problem with spurious singularities, that
we encountered in item 3~and~4 can be looked upon
as follows.
Formula~\ref{eq:split} tells us to split the invariant functions
in the matrix element. However, we have some freedom in making this
split-up. This makes it possible that the pole
term has a singularity that is then canceled by the regular term.
A sensible choice of such a split-up takes care not to
introduce new singularities in the physical phase space.

\end{document}